\documentclass[epj]{svjour}
\usepackage{graphics}
\begin{document}
\title{Role of the $qqqq\bar{q}$ components in the
electromagnetic transition $\gamma^{*}N\rightarrow N^{*}(1535)$}
\author{C. S. An\inst{1,}\inst{2}
\and B. S. Zou\inst{1,}\inst{3}
}                     
%
%
\institute{Institute of High Energy Physics, CAS, P.O.Box 918(4),
Beijing 100049, China \and Helsinki Institute of Physics, POB 64,
00014 University of Helsinki, Finland \and Theoretical Physics
Center for Science Facilities, CAS, P.O.Box 918(4), Beijing 100049,
China}
\date{Received: date / Revised version: date}
%
\abstract{ The helicity amplitudes $A^{p}_{1/2}$ and $S^{p}_{1/2}$
for the electromagnetic transition $\gamma^{*} N\rightarrow
N^{*}(1535)$ are calculated in the quark model that is extended to
include the lowest lying $qqqq\bar{q}$ components in addition to the
$qqq$ component. It is  found that with admixture of 5-quark
components with a proportion of 20\% in the nucleon and 25-65\% in
the $N^{*}(1535)$ resonance the calculated helicity amplitude
$A^{p}_{1/2}$ decreases at the photon point, $Q^{2}=0$ to the
empirical range. The $qqqq\bar{q}$ components contain $s\bar s$
pairs, which is consistent with the substantial width for $N\eta$
decay of the $N^{*}(1535)$. The best description of the momentum
dependence of the empirical helicity amplitudes is obtained by
assuming that the $qqqq\bar{q}$ components are more compact than the
$qqq$ component. However, this version of extended quark model still
does not lead to a satisfactory simultaneous description of both
$A^{p}_{1/2}$ and $S^{p}_{1/2}$ although with significant
improvement.
} 
\maketitle
\section{Introduction}
\label{sec:1}

The structure of the nucleon resonances with spin-parity $1/2^-$ has
continued to be somewhat enigmatic. On the one hand the number of
observed resonances coincides with that predicted by the constituent
quark model with a monotonic confining interaction
\cite{isgur,glozman}, while on the other hand several of these
resonances can be dynamically generated in chiral meson-nucleon
models \cite{weise,oset}. The question then is whether a 3 quark or
a 4 quark + 1 antiquark description is the more appropriate one. The
$N^{*}(1535)$ resonance is particularly interesting in this regard,
as it has a sizeable $N\eta$ decay branch, even though its energy is
very close to the threshold of that decay. The electromagnetic
transition form factors of this resonance may in fact contain
crucial information to settle this issue. It has recently been shown
that the chiral dynamical approach can provide a description of both
the $A_{1/2}^p$ and the $S_{1/2}^p$ transition form factors
\cite{oset2}, while a simultaneous description of both these form
factors has proven elusive in the basic $qqq$ constituent quark
model \cite{capstick}.

To bridge the gap between the constituent quark model and the chiral
meson-nucleon resonance models, it is natural to extend the former
to include explicit $qqqq\bar{q}$ components in addition to its
basic $qqq$ structure. In the case of the $\Delta(1232)$ and the
$N^{*}(1440)$ resonances it has been shown that already a modest
admixture of $qqqq\bar q$ components can remove the main
under-prediction of the decay widths that is typical of the $qqq$
quark model \cite{riska,riska1,riska2}. Moreover, it has recently
been shown that the coupling of $N^{*}(1535)N\phi$ may be
significant \cite{xie}, which is consistent with the previous
indications of the notable $N^{*}(1535)K\Lambda$ coupling
\cite{zou2}. These suggest that there are large $s\bar{s}$
components in the $N^{*}(1535)$ resonance. With this suggestion, we
can give a natural explanation of the mass ordering of the
$N^{*}(1440)1/2^{+}$, $N^{*}(1535)1/2^{-}$ and
$\Lambda^{*}(1405)1/2^{-}$ resonances \cite{zou2}.

Here we calculate the helicity transition amplitudes $A_{1/2}^{p}$
and $S_{1/2}^{p}$ for the electromagnetic transition $\gamma^{*}p\to
N^{*}(1535)$, by including contributions from both $qqq$ components
and those $qqqq\bar{q}$ components with configurations expected to
have the lowest energy in the proton and $N^{*}(1535)$. These
amplitudes are not well described in the conventional $qqq$
constituent quark model \cite{qqq,qqq1}. The results show that the
calculated transition amplitudes can be improved by taking inclusion
of the $qqqq\bar{q}$ components.

We take the flavor-spin configurations of the four-quark subsystem
in the $qqqq\bar{q}$ components in the proton to be
$[4]_{FS}[22]_F[22]_S$, which is likely to have the lowest energy
\cite{helminen}. In the case of the negative parity $N^{*}(1535)$
resonance the corresponding most likely lowest energy configuration
is $[31]_{FS}[211]_F[22]_S$ \cite{helminen}. The flavor
configuration $[211]_{F}$ in the latter requires that the
$qqqq\bar{q}$ component can only be $qqqs\bar{s}$. This implies a
large hidden strangeness in the $N^{*}(1535)$ resonance and strong
couplings to $N\phi$ and $K\Lambda$ states, which is consistent with
the results in Refs. \cite{xie} and \cite{zou2}. This feature is
also in line with the mechanism of dynamical resonance formation
within the coupled channel approach based on chiral $SU(3)$ with the
$K\Sigma$ quasi-bound state explanation for the
$N^{*}(1535)$~\cite{weise,oset,oset2,Kaiser,oset3}. For completeness
we also consider the contributions of $d\bar{d}$ or $u\bar{u}$
components in the $N^{*}(1535)$ and consequently, which appear in
the next-to-next-to-lowest-energy configuration
$[31]_{FS}[22]_{F}[31]_{S}$.

Some of the empirical results on the strangeness magnetic form
factors may be described, at least qualitatively, by $uuds\bar s$
configurations in the proton, where the $\bar s$ antiquark is in the
$S-$state \cite{zou,zou1}. Since configurations with the antiquark
in the $S-$state cannot be represented by long range pion or kaon
loop fluctuations, this motivates to a systematic extension of the
$qqq$ quark model to include the $qqqq\bar q$ configurations
explicitly. On the other hand, the overall descriptions of the
baryon magnetic moments may be improved by taking the $qqqq\bar{q}$
components contributions into account \cite{me,me1}.

This present manuscript is organized in the following way. The wave
functions of the proton and $N^{*}(1535)$ are given in section
\ref{sec:2}. The helicity amplitudes $A^{p}_{1/2}$ and $S^{p}_{1/2}$
for $\gamma^{*}p\rightarrow N^{*}(1535)$ are calculated in section
\ref{sec:3}. Finally section \ref{sec:4} contains a concluding
discussion.

\section{The wave functions of nucleon and $N^{*}(1535)$}
\label{sec:2}

 The internal nucleon wave functions that include $qqqq\bar q$
components in addition to the conventional $qqq$ components for the
proton and $N^{*}(1535)$ may be written in the following form:
\begin{eqnarray}
|P, s_{z}\rangle&=&A_{(P)3q}|qqq\rangle
+A_{(P)5q}\sum_{i}A_{i}|qqqq_{i}
\bar{q_{i}}\rangle\, ,\nonumber \\
|N^{*}(1535), s_{z}\rangle&=&A_{(N^{*})3q}|qqq\rangle
+A_{(N^{*})5q}\\\nonumber &&\times\sum_{i}A_{i}|qqqq_{i}
\bar{q_{i}}\rangle\, . \label{wave}
\end{eqnarray}
Here $A_{(P)3q}$ and $A_{(P)5q}$ are the amplitudes for the 3-quark
and 5-quark components in the nucleon, respectively; $A_{(N^{*})3q}$
and $A_{(N^{*})5q}$ are the corresponding factors for $N^{*}(1535)$.
The sum over i runs over all the possible $qqqq_{i}\bar{q_{i}}$
components, and the factors $A_{i}$ denotes the corresponding
coefficient for the $qqqq_{i}\bar{q}_{i}$ component. We give the
explicit forms of the $qqq$ and $qqqq\bar{q}$ components in the
proton and $N^{*}(1535)$ in the next two subsections.

\subsection{The wave functions of $qqq$ components}
\label{ssec1}

Here we take the wave functions for the $qqq$ components in proton
and $N^{*}(1535)$ to be the conventional ones, which can be
expressed in the following way in the harmonic oscillator quark
model:
\begin{eqnarray}
|P,s_{z}\rangle_{3q}
&=&\frac{1}{\sqrt{2}}(|\frac{1}{2},t_{z}\rangle_{+}
|\frac{1}{2},s_{z}\rangle_{+}+|\frac{1}{2},t_{z}\rangle_{-}
|\frac{1}{2},s_{z}\rangle_{-})\nonumber\\
&&\times\phi_{00}(\vec{\kappa}_{2})
\phi_{00}(\vec{\kappa}_{1})\, ,\nonumber\\
|N^{*}(1535),s_{z}\rangle_{3q}
&=&\frac{1}{2}\sum_{ms}C^{\frac{1}{2},s_{z}}_{1m,\frac{1}{2},s}
\{\phi_{1m}(\vec{\kappa}_{2})\phi_{00}(\vec{\kappa}_{1})
[|\frac{1}{2},t_{z}\rangle_{+}
|\frac{1}{2},s\rangle_{+}\nonumber\\
&&-|\frac{1}{2},t_{z}\rangle_{-}
|\frac{1}{2},s\rangle_{-}]-\{\phi_{00}(\vec{\kappa}_{2})\phi_{1m}(\vec{\kappa}_{1})
[|\frac{1}{2},t_{z}\rangle_{+}\nonumber\\
&&\times|\frac{1}{2},s\rangle_{-}+|\frac{1}{2},t_{z}\rangle_{-}
|\frac{1}{2},s\rangle_{+}]\}\, .\label{3qwfc}
\end{eqnarray}
Here $|\frac{1}{2},s_{(z)}\rangle_{\pm}$ and
$|\frac{1}{2},t_{z}\rangle_{\pm}$, with $t_{z}$ being the
z-component of the isospin, are spin and isospin wave functions of
mixed symmetry $[21]_{S}$ and $[21]_{F}$, respectively, in which '+'
denotes a state that is symmetric and '-' denotes a state that is
anti-symmetric under exchange of the spin or isospin of the first
two quarks. The momenta $\vec{\kappa}_{i}$ are defined by the three
quarks momenta as
\begin{eqnarray}
\vec{\kappa}_{1}&=&\frac{1}{\sqrt{2}}(\vec{p}_{1}-\vec{p}_{2})\, ,\nonumber\\
\vec{\kappa}_{2}&=&\frac{1}{\sqrt{6}}(\vec{p}_{1}+\vec{p}_{2}-2\vec{p}_{3})\, .\nonumber\\
\end{eqnarray}
The harmonic oscillator wave functions are
\begin{eqnarray}
\phi_{00}(\vec \kappa;\omega_{3})&=&{1\over (\omega_{3}^2\pi)^{3/4}}
\exp\{-\frac{\kappa^{2}}{2\omega_{3}^{2}}\}\, ,\\\
\phi_{1,\pm1}(\vec
\kappa;\omega_{3})&=&\mp\sqrt{2}\,{\kappa_{\pm}\over
\omega_{3}}\, \phi_{00}(\vec \kappa;\omega_{3})\, ,\\\
\phi_{10}(\vec \kappa;\omega_{3})&=&\sqrt{2}\,{\kappa_{0}\over
\omega_{3}}\, \phi_{00}(\vec \kappa;\omega_{3})\, .\
\end{eqnarray}
Here $\kappa_{\pm}\equiv\frac{1}{\sqrt{2}}(\kappa_{x}\pm
i\kappa_{y})$, and $\kappa_{0}\equiv\kappa_{z}$. The subscripts
denote the quantum numbers ($lm$) of the oscillator wave functions.

\subsection{The wave functions of $qqqq\bar{q}$
components} \label{ssec2}

If the hyperfine interaction between the quarks depends on spin and
flavor \cite{glozman}, the $qqqq$ subsystems of the $qqqq\bar{q}$
components with the mixed spatial symmetry $[31]_{X}$ are expected
to be the configurations with the lowest energy, and therefore most
likely to form appreciable components of the baryons with positive
parity. Consequently the flavor-spin state of the $qqqq$ subsystem
is most likely totally symmetric: $[4]_{FS}$. Moreover in the case
of the proton, the flavor-spin configuration of the four quark
subsystem $[4]_{FS}[22]_{F}[22]_{S}$, with one quark in its first
orbitally excited state, and the anti-quark in its ground state, is
likely to have the lowest energy \cite{helminen}, consequently, the
$qqqq_{i}\bar{q}_{i}$ components and the corresponding amplitudes in
the proton may be \cite{me}:
\begin{equation}
\sqrt{\frac{2}{3}}|[uudd]_{[22]_{F}}\bar{d}\rangle+\sqrt{\frac{1}{3}}
|[uuds]_{[22]_{F}}\bar{s}\rangle\, .\
\end{equation}

The explicit form of the wave functions for these two $q\bar{q}$
components may be expressed as \cite{zou1,me,riska2}:
\begin{eqnarray}
|p,s_{z}\rangle_{5q}&=&\frac{1}{\sqrt{2}}\sum_{a,b}\sum_{m,s}C^{\frac{1}{2}s_{z}}
_{1m,\frac{1}{2}s}C^{[1^{4}]}_{[31]_{a}[211]_{a}}[211]_{C}(a)[31]_{X,m}(a)\nonumber\\
&&\times[22]_{F}(b)
[22]_{S}(b)\bar{\chi}_{s}\psi({\vec{\kappa_{i}}})\, .\ \label{pwfc}
\end{eqnarray}
Here the flavor-spin, color and orbital wave functions of the $qqqq$
subsystem have been denoted by the Young patterns, respectively. The
sum over $a$ runs over the three configurations of the $[31]_{X}$
and $[211]_{C}$ representations of $S_{4}$ permutation group, and
the sum over $b$ runs over the two configurations of the $[22]_{F}$
and $[22]_{S}$ representations of $S_{4}$ permutation group,
respectively. The factors $C^{[1^{4}]}_{[31]_{a}[211]_{a}}$ denotes
the $S_{4}$ Clebsch-Gordan coefficients.

The explicit forms of the flavor-spin configurations have been given
in Ref. \cite{zou1}, and the explicit color-space part of the wave
function (\ref{pwfc}) may be expressed in the form \cite{riska2}:
\begin{eqnarray}
\psi_{C}(\{\vec{\kappa}_i\})&=&\frac{1}{\sqrt{3}}\{[211]_{C1}
\varphi_{1m}(\vec{\kappa}_{1})\varphi_{00}(\vec{\kappa}_{2})
\varphi_{00}(\vec{\kappa}_{3})
-[211]_{C2}\nonumber\\
&&\times\varphi_{00}(\vec{\kappa}_{1})\varphi_{1m}(\vec{\kappa}_{2})
\varphi_{00}(\vec{\kappa}_{3})+[211]_{C3}\nonumber\\
&&\times\varphi_{00}(\vec{\kappa}_{1})\varphi_{00}(\vec{\kappa}_{2})
\varphi_{1m}(\vec{\kappa}_{3})\}\varphi_{00}(\vec{\kappa}_{4})\, .\
\label{CX}
\end{eqnarray}

Here the two additional Jacobi momenta $\vec{\kappa}_{3}$ and
$\vec{\kappa}_{4}$ are defined as
\begin{eqnarray}
\vec{\kappa}_{3}&=&\frac{1}{\sqrt{12}}(\vec{p}_{1}+\vec{p}_{2}
+\vec{p}_{3}-3\vec{p}_{4})\, ,\\\
\vec{\kappa}_{4}&=&\frac{1}{\sqrt{20}}(\vec{p}_{1}+\vec{p}_{2}
+\vec{p}_{3}+\vec{p}_{4}-4\vec{p}_{5})\, .\
\end{eqnarray}
and the harmonic oscillator wave functions are
\begin{eqnarray}
\varphi_{00}(\vec \kappa;\omega_{5})&=&{1\over
(\omega_{5}^2\pi)^{3/4}}
\exp\{-\frac{\kappa^{2}}{2\omega_{5}^{2}}\}\, ,\\\
\varphi_{1,\pm1}(\vec
\kappa;\omega_{5})&=&\mp\sqrt{2}\,{\kappa_{\pm}\over
\omega_{5}}\, \varphi_{00}(\vec \kappa;\omega_{5})\, ,\\\
\varphi_{10}(\vec \kappa;\omega_{5})&=&\sqrt{2}\,{\kappa_{0}\over
\omega_{5}}\, \varphi_{00}(\vec \kappa;\omega_{5})\, .\
\end{eqnarray}

In the case of the wave functions for the $qqqq\bar{q}$ components
in the $S_{11}$ resonance $N^{*}(1535)$, the parity of which is
negative, are definitely not same as that of the proton. Negative
parity demands that all of the quarks and the anti-quark be in their
ground state, or states with higher even angular momentum. If the
flavor- and spin-dependent hyperfine interaction between the quarks
are employed, for the $qqqq$ subsystem, the flavor-spin
configuration $[31]_{FS}[211]_{F}[22]_{S}$ is expected to have the
lowest energy, and the orbital state should be completely symmetric
state $[4]_{X}$ \cite{helminen}. Consequently, the wave function of
the $qqqq\bar{q}$ component in $N^{*}(1535)$ may be expressed as
\begin{eqnarray}
|N^{*}(1535),s_{z}\rangle_{5q}&=&\sum_{abc}C^{[1^4]}_{[31]_{a}
[211]_{a}}C^{[31]_{a}}_{[211]_{b}[22]_{c}}[4]_{X}[211]_{F}(b)\nonumber\\
&&\times[22]_{S}(c)[211]_{C}(a) \bar \chi_{s_{z}}\varphi(\{\vec
\kappa_i\})\, \label{1535wfc}.
\end{eqnarray}

Note that the total spin of the four quark subsystem with symmetry
$[22]_{S}$ is $S=0$. The flavor configuration $[211]_{F}$ implies
that the $qqqq\bar{q}$ components can only be $uuds\bar{s}$.
Consequently there are appreciable $s\bar{s}$ components in
$N^{*}(1535)$, which is consistent with the strong coupling of
$N^{*}(1535)K\Lambda$ \cite{zou2} and $N^{*}(1535)N\phi$ \cite{xie}.
From this assumption it follows that the proportion of the
$s\bar{s}$ in the $N^{*}(1535)$ resonance might equal $P_{5q}$. In
the Roper resonance, the probability of the $s\bar{s}$ component
would then be only $P_{5q}/3$ \cite{me}. If we set same probability
$P_{5q}$ of the $qqqq\bar{q}$ components in $N^{*}(1535)$ and
$N^{*}(1440)$ the larger proportion of the strange component may
then make $N^{*}(1535)$ heavier than the Roper resonance as desired.
This has been a puzzle in the conventional $qqq$ constituent quark
model with flavor independent hyperfine interactions.

In addition, there may be smaller $d\bar{d}$ or $u\bar{u}$
components in $N^{*}(1535)$. The next-to-lowest-energy $qqqq\bar{q}$
configuration is however in this scheme
$[4]_{X}[31]_{FS}[211]_{F}[31]_{S}$ \cite{helminen}, which also
rules out the $u\bar{u}$ and $d\bar{d}$ components. The
contributions to the electromagnetic transition
$\gamma^{*}p\rightarrow N^{*}(1535)$ of this configuration may be
replaced by that of certain proportion of the lowest energy
$qqqq\bar{q}$ component. Consequently, one has to consider the
configuration with $qqqq$ subsystem of the orbital-flavor-spin
symmetry $[4]_{X}[31]_{FS}[22]_{F}[31]_{S}$, which is the lowest
energy configuration allowing $uudd\bar d$ component with
spin-parity $J^P=1/2^-$ \cite{helminen}. For this configuration, the
$qqqq_{i}\bar{q}_{i}$ components and the amplitudes in the
$N^{*}(1535)$ may be \cite{me}:
\begin{equation}
\sqrt{\frac{2}{3}}|[uudd]_{[22]_{F}}\bar{d}\rangle+\sqrt{\frac{1}{3}}
|[uuds]_{[22]_{F}}\bar{s}\rangle\, .\
\end{equation}
The explicit form of the wave function for these two $q\bar{q}$
components may be expressed as:
\begin{eqnarray}
|N^{*}(1535),s_{z}\rangle_{5q'}&=&\sum_{abc}C^{[1^4]}_{[31]_{a}
[211]_{a}}C^{[31]_{a}}_{[211]_{b}[22]_{c}}[4]_{X}[22]_{F}(b)\nonumber\\
&&\times[31]_{S}(c)[211]_{C}(a) \bar \chi_{s_{z}}\varphi(\{\vec
\kappa_i\})\, \label{1535wfc'}.
\end{eqnarray}

Actually, we can also obtain the probabilities of the $qqqq\bar{q}$
components from a recent manuscript \cite{charge}. The vanishing or
small axial charge of $N^{*}(1535)$ requires that there should be a
cancelation between the contributions of the $qqq$ and $qqqq\bar{q}$
components. Consequently, the most obvious five-quarks components
should be the ones in which the flavor-spin configurations of the
four-quarks subsystem are $[31]_{FS}[211]_{F}[22]_{S}$,
$[31]_{FS}[211]_{F}[31]_{S}$ or $[31]_{FS}[31]_{F}[31]_{S}$. And the
large coefficients $A_{n}$ obtained from the latter two
configurations indicate that the probabilities of these two
components should be smaller than the first one.

\section{The electromagnetic transition $\gamma^{*}N\rightarrow N^{*}(1535)$}
\label{sec:3}

The calculation of the helicity amplitudes $A_{1/2}^{p}$ and
$S_{1/2}^{p}$ for the electromagnetic transition
$\gamma^{*}p\rightarrow N^{*}(1535)$ that takes into account the
contributions of the $qqqq\bar{q}$ components discussed above falls
into three parts: (1) the contributions of the direct transition
$\gamma^{*}qqq\rightarrow qqq$, i. e. the contributions of the $qqq$
components in proton and $N^{*}(1535)$, (2) the contributions of the
lowest energy $qqqq\bar{q}$ component in $N^{*}(1535)$, and (3) the
contributions of the next-to-next-to-lowest-energy $qqqq\bar{q}$
component in $N^{*}(1535)$. And each of the latter two parts
contains contributions from two processes: the direct transition
$\gamma^{*}qqqq\bar{q}\rightarrow qqqq\bar{q}$, i.e. the diagonal
transitions, and the annihilation transition
$\gamma^{*}qqq\rightarrow qqqq\bar{q}$ (or
$\gamma^{*}qqqq\bar{q}\rightarrow qqq$), i.e. the off-diagonal
transitions.

In the non-relativistic approximation the elastic and annihilation
current operators are:
\begin{eqnarray}
\langle\vec{p}'|j^{0}|\vec{p}\rangle_{elas}&=&1\, ,\nonumber\\
\langle\vec{p}'|j^{0}|\vec{p}\rangle_{anni}&=&\frac{1}{2m}
[\vec{\sigma}\cdot(\vec{p}'+\vec{p})]\, ,\nonumber\\
\langle\vec{p}'|\vec{j}|\vec{p}\rangle_{elas}&=&\frac{\vec{p}'+\vec{p}}
{2m}+\frac{i}{2m}(\vec{\sigma}\times\vec{q})\, ,\nonumber\\
\langle\vec{p}'|\vec{j}|\vec{p}\rangle_{anni}&=&\vec{\sigma}\, ,\
\label{current}
\end{eqnarray}
respectively. Here we have set $\vec{q}=\vec{p}'-\vec{p}$.

\subsection{The contributions of the $qqq$ components to
the helicity amplitude $A^{P}_{1/2}$ and $S^{P}_{1/2}$}
\label{s3sec1}

For point-like quarks the electromagnetic transition operator for
elastic transitions between states with $n_{q}$ quarks is
\begin{eqnarray}
\hat{T}_{A}&=&-\sum^{n_{q}}_{i=1}\frac{e_{i}}{2m_{i}}[\sqrt{2}\hat{\sigma
}_{i+}k_{\gamma}+(p_{i+}'+p_{i+})]\,\nonumber\\
\hat{T}_{S}&=&\sum^{n_{q}}_{i=1}e_{i}\, .\ \label{od}
\end{eqnarray}
Here $\hat{T}_{A}$ and $\hat{T}_{S}$ are the corresponding operators
for the transverse and longitudinal helicity amplitudes, which are
obtained by coupling the current operators (\ref{current}) to the
transverse
($\vec{\epsilon}_{+1}=-\frac{1}{\sqrt{2}}(\hat{x}+i\hat{y})$ and
$\epsilon_{0}=0$) and longitudinal ($\vec{\epsilon}=0$ and
$\epsilon_{0}=1$) photon \cite{Lee}, respectively. And $e_{i}$ and
$m_{i}$ denote the electric charge and constituent mass of the quark
which absorbs the photon, respectively. $p_{i+}$ and $p'_{i+}$ are
defined as the initial and final momenta of the quark that absorbs
the photon:
\begin{equation}
p_{i+}^{(')}\equiv\frac{1}{\sqrt{2}}(p_{ix}^{(')}+ip_{iy}^{(')})\,
,\
\end{equation}
and $\hat{\sigma}_{i+}$ is defined as
\begin{equation}
\hat{\sigma}_{i+}\equiv
\frac{1}{2}(\hat{\sigma}_{ix}+i\hat{\sigma}_{iy})\, ,\
\end{equation} which raises the spin of the quark which absorbs the
photon. Here we consider a right-handed virtual-photon in the
operator $\hat{T}_{A}$, and the momentum of photon has been set to
be: $\vec{k}=(0,0,k_{\gamma})$ in the center-of-mass frame of the
final resonance $N^{*}(1535)$. It is related to the initial and
final momentum of the quark which absorbs the photon as
$k_{\gamma}=\vec{p}_{i}'-\vec{p}_{i}$, and the magnitude of
four-momentum transfer $Q=\sqrt{-k^{2}}$ as:
\begin{equation}
k_{\gamma}^{2}=Q^{2}+\frac{(M^{*2}-M_{N}^{2}-Q^{2})^{2}}{4M^{*2}}\,
,\
\end{equation}
where $M^{*}$ and $M_{N}$ denote the mass of $N^{*}(1535)$ resonance
and proton, respectively.

With the diagonal transition operators (\ref{od}), the helicity
amplitudes of the electromagnetic transition $\gamma^{*}p\rightarrow
N^{*}(1535)$, which include only the contributions from the $qqq$
components in the proton and the $N^{*}(1535)$, may be expressed as
\begin{eqnarray}
A_{1/2}^{p}&=&\frac{1}{\sqrt{2K_{\gamma}}}\langle N^{*}(1535),
\frac{1}{2},\frac{1}{2}|\hat{T_{A}}| p, \frac{1}{2},
-\frac{1}{2}\rangle\, , \nonumber\\
S_{1/2}^{p}&=&\frac{1}{\sqrt{2K_{\gamma}}}\langle N^{*}(1535),
\frac{1}{2},\frac{1}{2}|\hat{T_{S}}| p, \frac{1}{2},
\frac{1}{2}\rangle\, \ .
\end{eqnarray}
Here $K_{\gamma}$ is the real-photon three-momentum in the center of
mass frame of $N^{*}(1535)$.

With the wave functions (\ref{3qwfc}) and the operators (\ref{od}),
we can obtain the helicity amplitudes $A_{1/2}^{p}$ and
$S_{1/2}^{p}$ in the following form:
\begin{eqnarray}
A^{p(3q)}_{1/2}&=&A_{(p)_{3q}}A_{(N^{*})3q}\frac{1}{\sqrt{2K_{\gamma}}}
\frac{e}{2m}(\frac{k_{\gamma}^{2}}
{3\omega_{3}}+\frac{2\omega_{3}}{3})\exp\{-\frac{k_{\gamma}^{2}}
{6\omega_{3}^{2}}\}\,  \nonumber\\
S^{p(3q)}_{1/2}&=&-A_{(p)_{3q}}A_{(N^{*})3q}\frac{1}{\sqrt{2K_{\gamma}}}
\frac{e}{3\sqrt{2}}\frac{k_{\gamma}}
{\omega_{3}}\exp\{-\frac{k_{\gamma}^{2}} {6\omega_{3}^{2}}\}\, .\
\end{eqnarray}

\begin{figure}
\resizebox{0.5\textwidth}{!}{%
  \includegraphics{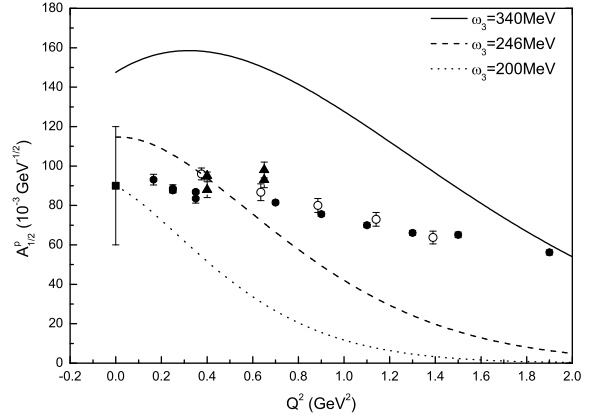}
}
\caption{The helicity amplitude $A^{p}_{1/2}$ for
$\gamma^{*}p\rightarrow N^{*}(1535)$ in the $qqq$ model. Here the
solid line is obtained by taking $\omega_{3}=340$ MeV, and the dash
and dot lines are obtained by taking $\omega_{3}=246$ MeV and
$\omega_{3}=200$ MeV, respectively. The data point at $Q^{2}=0$
(square) is from Ref. \cite{pdg}, the other points are taken from
Ref. \cite{prc} (triangles), \cite{clas} (open circles) and
\cite{clasnew} (filled circles).}
\label{fig1}       
\end{figure}

\begin{figure}
\resizebox{0.5\textwidth}{!}{%
  \includegraphics{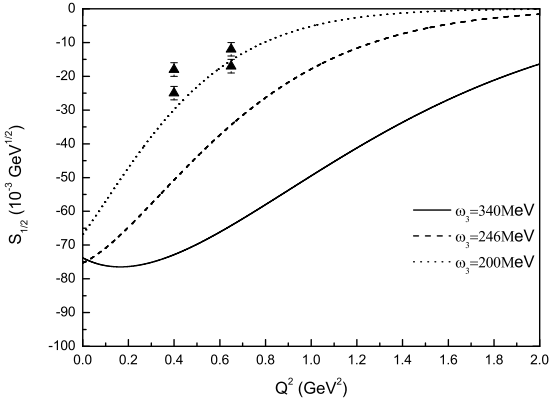}
}
\caption{The helicity amplitude $S^{p}_{1/2}$ for
$\gamma^{*}p\rightarrow N^{*}(1535)$ in the $qqq$ model. The data
points are extracted from Ref. \cite{prc}.}
\label{fig2}       
\end{figure}

Here $e$ denotes the electric charge of the proton. The model
parameters are the constituent masses of the light quarks $m$, the
oscillator parameter $\omega_{3}$, and the $qqq$ components
amplitudes $A_{3q}$. For the $qqq$ model, we set the constituent
quark mass to be $m=340$ MeV. The oscillator parameter $\omega_{3}$
may be determined by the nucleon radius as
$\omega_{3}=\frac{1}{\sqrt{\langle r^{2}\rangle}}$, which yield the
value $бл246$ MeV, while the empirical value for this parameter
falls in the range $110-410$ MeV \cite{riska2,Isgur2}. We give the
numerical results as functions of $Q^{2}$ by setting
$\omega_{3}=340$ MeV, $\omega_{3}=246$ MeV and $\omega_{3}=200$ MeV,
respectively, which are shown in figure \ref{fig1} and \ref{fig2}
for the transverse and longitudinal helicity amplitudes,
respectively, compared to the experimental data extracted from Ref.
\cite{pdg,prc,clas}. As in this case there are no $qqqq\bar{q}$
components in the proton and $N^{*}(1535)$ the amplitudes are
$A_{(p)3q}=A_{(N^{*})3q}=1$.

 As shown in figure \ref{fig1}, none of the three curves can
describe the experimental data satisfactorily. At the photon point,
$Q^{2}=0$, the calculated helicity amplitudes are
$A^{p}_{1/2}=0.147/\sqrt{\mathrm{GeV}}$ and
$A^{p}_{1/2}=0.115/\sqrt{\mathrm{GeV}}$ with the parameter
$\omega_{3}=340$ MeV and $\omega_{3}=246$ MeV, respectively, both of
which are larger than the experimental value
$A^{p}_{1/2}=0.090\pm0.030/\sqrt{\mathrm{GeV}}$ \cite{pdg}. For the
curve which is obtained by setting the parameter $\omega_{3}=200$
MeV, the calculated amplitude decribes the data better at the photon
point, but it falls too fast in comparison with the data when
$Q^{2}$ increases. Similarly in the case of the longitudinal
helicity amplitude $S^{p}_{1/2}$, none of the three curves can fit
the data well, as that all of them are too small near the photon
point.

\subsection{The contributions of the lowest energy $qqqq\bar{q}$ components of
$N^{*}(1535)$ to the helicity amplitude $A^{P}_{1/2}$ and
$S^{P}_{1/2}$} \label{s3sec2}

The contributions of the $qqqq\bar{q}$ components contain the direct
transition matrix elements $\gamma^{*}qqqq\bar{q}\rightarrow
qqqq\bar{q}$, and the annihilation transition elements
$\gamma^{*}qqq\rightarrow qqqq\bar{q}$ and
$\gamma^{*}qqqq\bar{q}\rightarrow qqq$, i. e. the diagonal and
non-diagonal transitions.

The contributions from the diagonal transition elements of the
$qqqq\bar{q}$ components are obtained as matrix elements of the
operator (\ref{od}) (with $n_{q}=5$) between the $5q$ wave functions
(\ref{pwfc}) and (\ref{1535wfc}), and between the wave functions
(\ref{pwfc}) and (\ref{1535wfc'}). At the first step, we consider
the contributions of the lowest energy $qqqq\bar{q}$ components in
$N^{*}(1535)$.

The lowest energy configuration in the $qqqq\bar{q}$ components in
$N^{*}(1535)$ is the one in which the flavor-spin configuration of
the four quark subsystem has the mixed symmetry
$[31]_{FS}[211]_{F}[22]_{S}$, as mentioned in section \ref{sec:2}.
For the spin symmetry $[22]_{S}$ ($S=0$), both in the proton and
$N^{*}(1535)$, the matrix elements of the operator
$\hat{\sigma}_{i+}$ of these four quarks vanish. On the other hand,
by the orthogonality of the four-quark orbital states $[31]_{X}$
(for the proton) and $[4]_{X}$ (for the $N^{*}(1535)$), the matrix
element of the operator $\hat{T}_{A}$ of the anti-quark does not
contribute to the transition $\gamma^{*}qqqq\bar{q}\rightarrow
qqqq\bar{q}$. Finally the contributions to $A^{p(5q)}_{1/2}$ only
come from the matrix element of the second term of the operator
$\hat{T}_{A}$ (\ref{od}) between the four quark states. In the case
of the helicity amplitude $S^{p(5q)}_{1/2}$, the matrix element of
the operator should vanish for the orthogonality of the different
$qqqq\bar{q}$ states in proton and $N^{*}(1535)$.

Explicit calculation yields:
\begin{eqnarray}
A^{p(5q)}_{1/2}&=&A_{(p)s\bar{s}}
A_{(N^{*})s\bar{s}}\frac{1}{\sqrt{2K_{\gamma}}}
\frac{\omega_{5}}{24}(\frac{e}{2m}-\frac{e}{2m_{s}})
\exp\{-\frac{k_{\gamma}^{2}}{5\omega_{5}^{2}}\}\, \nonumber\\
S^{p(5q)}_{1/2}&=&0 \, .\
\end{eqnarray}
Here we have neglected the transition between the $uudd\bar{d}$
component in the proton and the $uuds\bar{s}$ component in
$N^{*}(1535)$, which should be very tiny.

Consider the the contributions of the non-diagonal transition
elements, i. e. the transitions $\gamma^{*}qqq\rightarrow
qqqq\bar{q}$ and $\gamma^{*}qqqq\bar{q}\rightarrow qqq$. By equation
(\ref{current}), the operator for these transitions may be expressed
as
\begin{eqnarray}
\hat{T}_{Aanni}&=&-\sum^{4}_{i=1}\sqrt{2}e_{i}\hat{\sigma}_{i+}\,,\nonumber\\
\hat{T}_{Sanni}&=&\sum^{4}_{i=1}\frac{e_{i}}{2m_{i}}\vec{\sigma}
\cdot(\vec{p}_{i}+\vec{p}_{5}) \, .\label{nd}
\end{eqnarray}
Here $\hat{T}_{Aanni}$ and $\hat{T}_{Sanni}$ denote the transition
operators for $A^{p(anni)}_{1/2}$ and $S^{p(anni)}_{1/2}$,
respectively, and $e_{i}$ and $m_{i}$ are the electric charge and
the constituent mass of the annihilating quark, and
$\hat{\sigma}_{i+}$ raises the spin of the corresponding quark.
$\vec{p}_{i}$ is the momentum of the annihilating quark and
$\vec{p}_{5}$ the momentum of the antiquark.

First, we calculate the matrix elements for the transitions
$\gamma^{*}qqq\rightarrow qqqq\bar{q}$. These involve calculations
of the overlap between the $\gamma^{*}qqq$ wave function and that
for the $qqqq\bar{q}$ component in $N^{*}(1535)$.

In the case of the lowest energy configurations, the flavor-spin
configuration has the mixed symmetry\\ $[31]_{FS}[211]_{F}[22]_{S}$,
the explicit form of which is shown in appendix \ref{seca}. For the
case of the color overlap, the only contribution comes from the
color symmetry configuration of the $qqqq\bar{q}$ component in
$N^{*}(1535)$ which is denoted by $[211]_{C3}$. The matrix elements
between the color singlet of the $qqq$ component in the proton and
the other two color configurations ($[211]_{C1}$ and $[211]_{C2}$)
vanish.

Explicit calculation leads to the result:
\begin{eqnarray}
A_{1/2}^{p(3q\rightarrow5q)}&=&-A_{(p)3}A_{(N^{*})s\bar{s}}
\frac{1}{\sqrt{2K_{\gamma}}}\frac{\sqrt{2}}{6}
eC_{35}\exp\{-\frac{3k_{\gamma}^{2}}{20\omega_{5}^{2}}\}\,,\nonumber\\
S_{1/2}^{p(3q\rightarrow5q)}&=&A_{(p)3}A_{(N^{*})s\bar{s}}
\frac{1}{\sqrt{2K_{\gamma}}}\frac{1}{6} \frac{e}{2m_{s}}k_\gamma
C_{35}\nonumber\\
&&\times\exp\{-\frac{3k_{\gamma}^{2}}{20\omega_{5}^{2}}\}\,.
\label{anni}
\end{eqnarray}
Here the factor $C_{35}$ denotes the orbital overlap factor:
\begin{equation}
\langle\varphi_{00}(\vec{\kappa}_{1})\varphi_{00}(\vec{\kappa}_{2})
|\varphi_{00}(\vec{\kappa}_{1})\varphi_{00}(\vec{\kappa}_{2})\rangle\
=(\frac{2\omega_{3}\omega_{5}}{\omega_{3}^{2}+\omega_{5}^{2}})^{3}
\, .
\end{equation}

\begin{figure}
\resizebox{0.5\textwidth}{!}{%
  \includegraphics{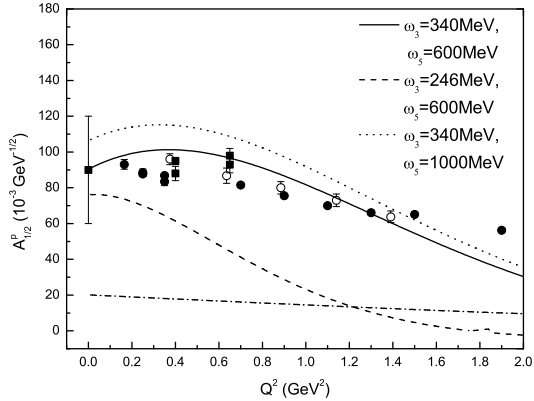}
}
\caption{The helicity amplitude $A^{p}_{1/2}$ for
$\gamma^{*}p\rightarrow N^{*}(1535)$ contains the contributions both
of the $qqq$ and lowest energy $qqqq\bar{q}$ components in the
proton and $N^{*}(1535)$. Here the solid line is obtained by taking
$\omega_{3}=340$ MeV and $\omega_{5}=600$ MeV, the dash line
$\omega_{3}=246$ MeV and $\omega_{5}=600$ MeV, and the dot line
$\omega_{3}=340$ MeV and $\omega_{5}=1000$ MeV, respectively. And
the dash dot line is the absolute value of the contributions of the
5-quark component with $\omega_{3}=340$ and $\omega_{5}=600$ MeV.
Data point as in figure \ref{fig1}.}
\label{fig3}       
\end{figure}

\begin{figure}
\resizebox{0.5\textwidth}{!}{%
  \includegraphics{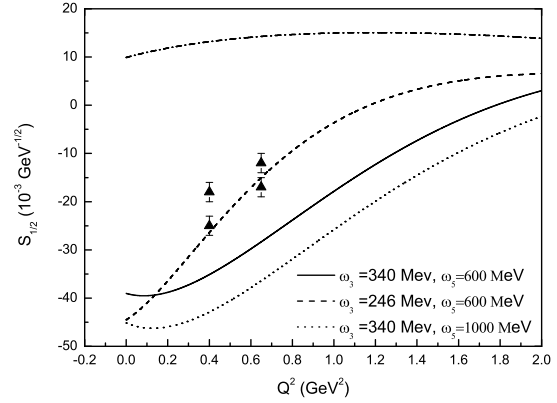}
}
\caption{The helicity amplitude $S^{p}_{1/2}$ for
$\gamma^{*}p\rightarrow N^{*}(1535)$ contains the contributions both
of the $qqq$ and lowest energy $qqqq\bar{q}$ components in the
proton and $N^{*}(1535)$.  The dash dot line is the emperical value
of the contributions of the 5-quark component with $\omega_{3}=340$
and $\omega_{5}=600$ MeV. Data point as in figure \ref{fig2}.}
\label{fig4}       
\end{figure}

Here we have obtained the helicity amplitudes $A^{p}_{1/2}$ and
$S^{p}_{1/2}$ for the electromagnetic transition
$\gamma^{*}p\rightarrow N^{*}(1535)$, which contains the
contributions both of the $qqq$ and the lowest energy $qqqq\bar{q}$
components in the proton and $N^{*}(1535)$. The results are shown in
figure \ref{fig3} and \ref{fig4}, respectively. Here we have taken
the probability of the $qqqq\bar{q}$ components in the proton as the
tentative value $P_{5q}=20\%$, and in $N^{*}(1535)$ $P_{5q}=45\%$.
Taking the $qqqq\bar{q}$ components into account, the constituent
quarks masses should be a bit smaller than the ones we employ in the
previous section. To reproduce the mass for the nucleon when the
five-quark components have been included, we take the values
$m_{u}=m_{d}=m=290$ MeV, and $m_{s}=430$ MeV.  The oscillator
parameters are $\omega_{3}$ and $\omega_{5}$. The latter one is
treated as a free parameter in this manuscript. Note that the value
for the $\omega_{5}$ may be different from that for $\omega_{3}$. In
our extended quark model with each baryon as a mixture of the
three-quark and five-quark components, the two components represent
two different states of the baryon. For the $qqqq\bar{q}$ state,
there are more color sources than the $qqq$ state, and may make the
effective phenomenological confinement potential stronger. This is
consistent with other empirical evidence favoring larger value of
the $\omega_{5}$~\cite{riska,riska1,riska2,me}. An intuitive picture
for our extended quark model is like this: the $qqq$ state has
weaker potential; when quarks expand, a $q\bar{q}$ pair is pulled
out and results in a $qqqq\bar{q}$ state with stronger potential;
the stronger potential leads to a more compact state which then
makes the $\bar{q}$ to annihilate with a quark easily and transits
to the qqq state; this leads to constantly transitions between these
two states. Here presented are the results by setting
$\omega_{3}=340$ MeV and $\omega_{5}=600$ MeV, $\omega_{3}=246$  and
$\omega_{5}=600$ MeV, and $\omega_{3}=340$ MeV and $\omega_{5}=1000$
MeV, respectively. As shown in figure \ref{fig3}, the results
describe the experimental data for $A^{p}_{1/2}$ well when the
oscillator parameters are given the values $\omega_{3}=340$ MeV and
$\omega_{5}=600$ MeV, both at the photon point and larger $Q^{2}$.
While in figure \ref{fig4}, the magnitude of the values for
$S^{p}_{1/2}$ are however larger than the experimental value, even
though the momentum dependence appears reasonable. Intriguingly when
$Q^{2}$ increases to about $1.8 GeV^{2}$ the calculated values
change sign. So here we should conclude that this model can work
when the $Q^{2}$ is less than $1.8 GeV^{2}$, if $Q^{2}$ is larger
than this value, maybe we should consider the relativistic effect.

There are another important parameter in our model, the phase
factors $\delta$ between the $qqq$ and $qqqq\bar{q}$ components of
the $N^{*}(1535)$ resonance, which has been taken to be $+1$. But in
principle, this factor could be an arbitrary complex one
$\exp\{i\phi\}$. As we know, the helicity amplitude $A^{p}_{1/2}$ is
real, so here we may choose $\delta$ to be $\pm1$. And as shown in
figure \ref{fig3}, the non-diagonal transition contributes a minus
value to the $A^{p}_{1/2}$. If we assume $\delta$ to be $-1$, the
numerical value for $A^{p}_{1/2}$ may be about
$0.110/\sqrt{\mathrm{GeV}}$, the result is not so good.
Consequently, the best choice for us may be $\delta=+1$.

Note that the diagonal contributions of the lowest energy
$qqqq\bar{q}$ component in $N^{*}(1535)$ to $S^{p}_{1/2}$ is $0$.
Actually, the contributions of the diagonal transition to
$A^{p}_{1/2}$ are also very small, which is less than
$0.01/\sqrt{\mathrm{GeV}}$. But the non-diagonal contributions are
significant, and negative for $A^{p}_{1/2}$ and positive for
$S^{p}_{1/2}$, which are shown in figure \ref{fig3} and \ref{fig4}
by the dash dot curves, with the parameter values $\omega_{3}=340$
MeV and $\omega_{5}=600$ MeV. For instance, at the photon point,
$Q^{2}=0$, with the parameter values $\omega_{3}=340$ MeV and
$\omega_{5}=600$ MeV, the contribution to $A^{p}_{1/2}$ is about
$-0.025/\sqrt{\mathrm{GeV}}$, which can decrease the helicity
amplitude $A^{p}_{1/2}$ to describe the data.

\subsection{The contributions of the next-to-next-to-lowest-energy
$qqqq\bar{q}$ components in $N^{*} (1535)$ to the helicity amplitude
$A^{P}_{1/2}$ and $S^{P}_{1/2}$} \label{s3sec3}

 The next-to-next-to-lowest-energy configuration of the \\$qqqq\bar{q}$
components in $N^{*}(1535)$ is the one in which the flavor-spin
configuration of the four quark subsystem has the mixed symmetry
$[31]_{FS}[22]_{F}[31]_{S}$, as mentioned in section \ref{sec:2}.
For the same reason as that in section \ref{s3sec2}, the matrix
element of the operator $\hat{T}_{A}$ between the anti-quark states
vanishes, and the diagonal contributions to $S^{p}_{1/2}$ is $0$.
The difference is, there are two $qqqq\bar{q}$ components in
$N^{*}(1535)$ with the four quark flavor-spin configuration
$[31]_{FS}[22]_{F}[31]_{S}$. Here we neglect the transitions
$\gamma^{*}uudd\bar{d}\rightarrow uuds\bar{s}$ and
$\gamma^{*}uuds\bar{s}\rightarrow uudd\bar{d}$. Calculation leads to
the result:
\begin{eqnarray}
A^{p(5q')}_{1/2}&=&-A_{(p)d\bar{d}}A_{(N^{*})d\bar{d}}\frac{1}{\sqrt{2K_{\gamma}}}
\frac{k_{\gamma}^{2}}{18\omega_{5}}\frac{e}{2m}
\exp\{-\frac{k_{\gamma}^{2}}{5\omega_{5}^{2}}\}-\nonumber\\
&&A_{(p)s\bar{s}}A^{'}_{(N^{*})s\bar{s}}\frac{1}{\sqrt{2K_{\gamma}}}
\frac{k_{\gamma}^{2}}{12\omega_{5}}
(\frac{e}{2m}-\frac{e}{6m_{s}})\nonumber\\
&&\times\exp\{-\frac{k_{\gamma}^{2}}{5\omega_{5}^{2}}\}\, ,\nonumber\\
S^{p5q'}_{1/2}&=&0\, .\
\end{eqnarray}

In the case of the annihilation transitions straightforward
calculation leads to the result that the flavor-spin overlap factors
for the transition $\gamma^{*}qqq\rightarrow qqqq\bar{q}$ $C_{FS}$
vanish, both for $A^{p}_{1/2}$ and $S^{p}_{1/2}$. Therefore this
transition does not contribute to the process
$\gamma^{*}p\rightarrow N^{*}(1535)$.

The next step is to consider the contributions of the transitions
$\gamma^{*}qqqq\bar{q}\rightarrow qqq$. The flavor-spin
configuration for the four-quarks subsystem of the $qqqq\bar{q}$
components in the proton is $[4]_{FS}[22]_{F}[22]_{S}$, for which
the explicit form has been given in Ref. \cite{zou1}. After some
calculation, it emerges that the flavor-spin overlap factors are
also $0$. Consequently, this process also yields no contribution to
the transition $\gamma^{*}p\rightarrow N^{*}(1535)$.

Finally, we find that the the next-to-next-to-lowest-energy
$qqqq\bar{q}$ components in $N^{*} (1535)$ does not contribute to
the helicity amplitude $S^{p}_{1/2}$ for the electromagnetic
transition $\gamma^{*}p\rightarrow N^{*}(1535)$. The contributions
to $A^{p}_{1/2}$ only comes from the diagonal transition elements.
As in the case of the contributions of the lowest energy
$qqqq\bar{q}$ components, this contribution is also very small. And
as the the non-diagonal contribution to $A^{p}_{1/2}$ is 0, this
5-quark component does not contribute significantly to the last
results. The result, which contains all the contributions for
$A^{p}_{1/2}$ that we have considered is shown in figure \ref{fig5}.
For comparison, we have considered several different probabilities
of the $qqqq\bar{q}$ components in $N^{*}(1535)$, as is discussed in
details in section \ref{s3sec5}.

\subsection{The helicity amplitude $A^{n}_{1/2}$ at the photon point}
\label{s3sec4}

As we know, the ratio of the helicity amplitudes at photon point for
proton and neutron is not a trivial issue. So we calculate the
$A^{n}_{1/2}$ in this section. For isospin symmetry, the wave
functions for neutron and $n^{*}(1535)$ can be obtained by the wave
functions we have given for the proton and $p^{*}(1535)$.
Consequently, we can calculate the matrix elements directly. After
some calculations, we can get the following results:
\begin{eqnarray}
A^{n(3q)}_{1/2}&=&-A_{(p)_{3q}}A_{(N^{*})3q}\frac{1}{\sqrt{2K_{\gamma}}}
\frac{e}{2m}(\frac{k_{2\gamma}^{2}}
{3\omega_{3}}+\frac{2\omega_{3}}{9})\exp\{-\frac{k_{\gamma}^{2}}
{6\omega_{3}^{2}}\}\,  \nonumber\\
A^{n(5q)}_{1/2}&=&-A_{(p)s\bar{s}}
A_{(N^{*})s\bar{s}}\frac{1}{\sqrt{2K_{\gamma}}}
\frac{\omega_{5}}{24}(\frac{e}{2m}+\frac{e}{2m_{s}})
\exp\{-\frac{k_{\gamma}^{2}}{5\omega_{5}^{2}}\}\,  \nonumber\\
A_{1/2}^{n(3q\rightarrow5q)}&=&-A_{(p)3}A_{(N^{*})s\bar{s}}
\frac{1}{\sqrt{2K_{\gamma}}}\frac{\sqrt{2}}{6}
eC_{35}\exp\{-\frac{3k_{\gamma}^{2}}{20\omega_{5}^{2}}\}\,.\
\label{neutron}
\end{eqnarray}
As we can see in equation (\ref{neutron}), the non-diagonal
contributions to the helicity amplitude $A^{n}_{1/2}$ is same as the
one to $A^{p}_{1/2}$. Note that here we have only considered the
contributions of the lowest energy five-quark components in
$n^{*}(1535)$, for that the probability of the
next-to-next-to-lowest-energy $qqqq\bar{q}$ components should be
some smaller, and on the other side, we can see in the section
\ref{s3sec3}, the non-diagonal contributions between the three-quark
component and this five-quark configuration is $0$, so it may only
give a very small correction to our result, so we can neglect it
here.

If we set $\omega_{3}=340$ MeV, and the constituent quark masses are
taken to be the same value as that in section \ref{s3sec1}, then we
can get that $A^{n}_{1/2}=-0.123/\sqrt{\mathrm{GeV}}$, which is much
larger than the data
$A^{n}_{1/2}=-0.046\pm0.027/\sqrt{\mathrm{GeV}}$ \cite{pdg}. It
indicates that we should consider the contributions of the
five-quark components. When the contributions of the $qqqq\bar{q}$
components are taken into account, with the same parameter employed
in section \ref{s3sec2} (That for the best fit), we can get that the
helicity amplitude $A^{n}_{1/2}=-0.074/\sqrt{\mathrm{GeV}}$, which
falls in the range of the data. And the ratio
$A^{n}_{1/2}/A^{p}_{1/2}$ is then $-0.82$, which also falls well in
the range of the data $0.84\pm0.15$. Note that the non-diagonal
contributions to $A^{n}_{1/2}$ is same as that to $A^{p}_{1/2}$, and
as we have obtained in section \ref{s3sec2} that this contribution
should be the major one of the $qqqq\bar{q}$ components, so we can
conclude that the ratio $A^{n}_{1/2}/A^{p}_{1/2}$ is not sensitive
with the free parameter $\omega_{5}$.

\subsection{The probabilities of the $qqqq\bar{q}$ components in $N^{*}(1535)$}
\label{s3sec5}

\begin{figure}
\resizebox{0.5\textwidth}{!}{%
  \includegraphics{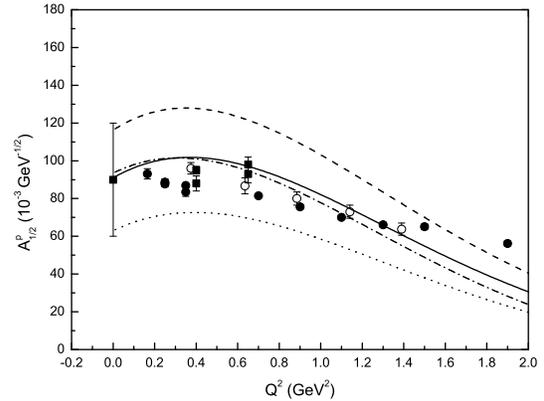}
}
\caption{The total helicity amplitude $A^{p}_{1/2}$ for
$\gamma^{*}p\rightarrow N^{*}(1535)$. Here all the lines are
obtained by taking $\omega_{3}=340$ MeV and $\omega_{5}=600$ MeV.
The solid line is obtained by taking the probability of the lowest
energy $qqqq\bar{q}$ components in $N^{*}(1535)$ to be the totally
$P_{5q}$, and the dash dot line $0.6P_{5q}$ MeV, i. e. the
probability of the next-to-next-to-lowest energy $qqqq\bar{q}$
components is $0.4P_{5q}$, and both the two lines are obtained by
setting $P_{3q}=0.55$. The dot line is obtained by setting
$P_{3q}=0.35$, and the dash line $P_{3q}=0.75$, and both of the two
lines are obtained by taking the probability of the lowest energy
$qqqq\bar{q}$ component to be the totally $P_{5q}$. Data point as in
figure \ref{fig1}.}
\label{fig5}       
\end{figure}

Here we have calculated the helicity amplitude $A^{p}_{1/2}$ by
considering all the contributions of the $qqq$ and low lying
$qqqq\bar{q}$ components in the proton and $N^{*}(1535)$. The last
result is shown in figure \ref{fig5} as a function of $Q^{2}$. Here
all the lines are obtained by taking $\omega_{3}=340$ MeV and
$\omega_{5}=600$ MeV. The solid line is obtained by taking the
probability of the lowest energy $qqqq\bar{q}$ components in
$N^{*}(1535)$ to be the totally $P_{5q}$, and the dash dot line
$0.6P_{5q}$ MeV, i.e. the probability of the
next-to-next-to-lowest-energy $qqqq\bar{q}$ components is
$0.4P_{5q}$, and both the two lines are obtained by setting
$P_{3q}=55\%$. The dot line is obtained by setting $P_{3q}=35\%$,
and the dash line $P_{3q}=75\%$, and both of the two lines are
obtained by taking the probability of the lowest energy
$qqqq\bar{q}$ component to be the totally $P_{5q}$.

As shown in figure \ref{fig5}, the best description of $A^{p}_{1/2}$
is given by the curve obtained by taking the probability of the
lowest energy $qqqq\bar{q}$ components in $N^{*}(1535)$ to be
$P_{5q}=45\%$. This indicates that this electromagnetic transition
does not favor any large probability of the next-to-next-to-lowest
energy $qqqq\bar{q}$ components in $N^{*}(1535)$, although its
contribution is tiny and the result is not very sensitive to it. The
main result is that if the probability of the lowest energy $5q$
components falls in the range  25-65\% the calculated helicity
amplitudes may fall within the data range $90\pm30/\sqrt{GeV}$ at
the photon point.

\section{Conclusion}
\label{sec:4}

The helicity amplitudes $A^{p}_{1/2}$ and $S^{p}_{1/2}$ for the
electromagnetic transition $\gamma^{*}p\rightarrow N^{*}(1535)$ were
calculated, and the possible role of the $qqqq\bar{q}$ was
investigated. The results indicate that the extension of the $qqq$
quark model to include $qqqq\bar q$ components can bring about a
much better description of the empirical results. The results
indicate that the admixtures of $qqqq\bar{q}$ components in the
$N^{*}(1535)$ resonance might be in the range $25-65\%$, and about
$20\%$ in the proton.

The orbital-flavor-spin configuration of the four-quark subsystem in
the $qqqq\bar{q}$ components of the proton was assumed to be
$[31]_{X}[4]_{FS}[22]_{F}[22]_{S}$, a configuration which is
expected to have the lowest energy, and therefore most likely to
form appreciable components of the proton. In the case of the
resonance $N^{*}(1535)$, the lowest energy configuration is
$[4]_{X}[31]_{FS}[211]_{F}[22]_{S}$, which requires that the $5q$
component of $N^{*}(1535)$ should only be $s\bar{s}$ component. This
means that there may be large strangeness components in
$N^{*}(1535)$ resonance, which is consistent with the strong
couplings of $N^{*}(1535)N\phi$ and $N^{*}(1535)K\Lambda$, predicted
in Refs. \cite{xie} and \cite{zou2}. This is also in line with the
$K\Sigma$ quasibound state explanation for $N^{*}(1535)$ by the
mechanism of dynamical resonance formation within the coupled
channel approach based on chiral $SU(3)$
\cite{weise,oset,oset2,Kaiser}. In addition we considered the
contributions of the $uudd\bar d$ and $uudu\bar{u}$ components in
the $N^{*}(1535)$, which have the orbital-flavor-spin configuration
$[4]_{X}[31]_{FS}[22]_{F}[31]_{S}$. The results show that this
electromagnetic transition does not favor large $qqqq\bar{q}$
components with the next-to-next-to-lowest-energy.

The suggested large probability $s\bar{s}$ components in
$N^{*}(1535)$, may be naturally consistent with the mass ordering of
the resonances $N^{*}(1440)$ and $N^{*}(1535)$. For the Roper
resonance, the largest $5q$ component is $uudd\bar{d}$
\cite{riska2}, while that for $N^{*}(1535)$ is $uuds\bar{s}$
component, which may make it heavier than the roper resonance. If we
neglect higher energy configurations of the $qqqq\bar{q}$ components
in these two resonances, the $s\bar{s}$ components in the Roper
should be $P_{5q}/3$ \cite{me}, in the case of $N^{*}(1535)$, it is
$P_{5q}$, which may make it heavier than the Roper resonance.

\section{Acknowledgments}

We are grateful to Professor D. O. Riska for helpful discussions and
English editing of the draft, and Dr Q. B. Li for helpful
discussions. This work is partly supported by the National Natural
Science Foundation of China under grants Nos. 10435080, 10521003,
and by the Chinese Academy of Sciences under project
No.~KJCX3-SYW-N2.

\begin{appendix}

\section{The explicit form of the flavor-spin configuration $[31]_{FS}[211]_{F}[22]_{S}$}
\label{seca}

 The explicit decomposition of the flavor-spin
configuration $[31]_{FS}[211]_{F}[22]_{S}$ may be expressed as
\cite{chen}
\begin{eqnarray}
|[31]_{FS}\rangle_{1}&=&\frac{1}{2}\{\sqrt{2}|[211]\rangle_{F1}|[22]\rangle_{S1}-
|[211]\rangle_{F2}|[22]\rangle_{S1}\nonumber\\
&&+|[211]\rangle_{F3}|[22]\rangle_{S3}\}\,,\\
|[31]_{FS}\rangle_{2}&=&\frac{1}{2}\{\sqrt{2}|[211]\rangle_{F1}|[22]\rangle_{S2}+
|[211]\rangle_{F2}|[22]\rangle_{S2}\nonumber\\
&&+|[211]\rangle_{F3}|[22]\rangle_{S1}\}\,,\\
|[31]_{FS}\rangle_{3}&=&\frac{1}{\sqrt{2}}\{-|[211]\rangle_{F2}|[22]\rangle_{S2}\nonumber\\
&&+|[211]\rangle_{F3}|[22]\rangle_{S1}\}\,,
\end{eqnarray}
and the explicit forms of the flavor symmetry $[211]_{F}$ and spin
symmetry $[22]_{S}$ are
\begin{eqnarray}
|[211]\rangle_{F1}&=&\frac{1}{4}\{2|uuds\rangle-2|uusd\rangle-|duus\rangle-|udus\rangle\nonumber\\
&&-|sudu\rangle-|usdu\rangle+|suud\rangle+|dusu\rangle\nonumber\\
&&+|usud\rangle+|udsu\rangle\}\,,\\
|[211]\rangle_{F2}&=&\frac{1}{\sqrt{48}}\{3|udus\rangle
-3|duus\rangle+3|suud\rangle-3|usud\rangle\nonumber\\
&&+2|dsuu\rangle-2|sduu\rangle-|sudu\rangle+|usdu\rangle+\nonumber\\
&&|dusu\rangle-|udsu\rangle\}\,,\\
|[211]\rangle_{F3}&=&\frac{1}{\sqrt{6}}\{|sudu\rangle+|udsu\rangle
+|dsuu\rangle-|usdu\rangle\nonumber\\
&&-|dusu\rangle-|sduu\rangle\}\,,
\end{eqnarray}
\begin{eqnarray}
|[22]\rangle_{S1}&=&\frac{1}{\sqrt{12}}\{2
|\uparrow\uparrow\downarrow\downarrow\rangle
+2|\downarrow\downarrow\uparrow\uparrow\rangle
-|\downarrow\uparrow\uparrow\downarrow\rangle\nonumber\\
&&-|\uparrow\downarrow\uparrow\downarrow\rangle
-|\downarrow\uparrow\downarrow\uparrow\rangle
-|\uparrow\downarrow\downarrow\uparrow\rangle\}\,,\\
|[22]\rangle_{S2}&=&\frac{1}{2}\{
|\uparrow\downarrow\uparrow\downarrow\rangle
+|\downarrow\uparrow\downarrow\uparrow\rangle
-|\downarrow\uparrow\uparrow\downarrow\rangle
-|\uparrow\downarrow\downarrow\uparrow\rangle\}\,.
\end{eqnarray}

\section{The explicit form of the flavor-spin configuration $[31]_{FS}[22]_{F}[31]_{S}$}
\label{secb}

 The explicit decomposition of the flavor-spin
configuration $[31]_{FS}[22]_{F}[31]_{S}$ may be expressed as
\cite{chen}
\begin{eqnarray}
|[31]_{FS}\rangle_{1}&=&\frac{1}{\sqrt{2}}\{|[22]\rangle_{F1}|[31]\rangle_{S2}+
|[22]\rangle_{F2}|[31]\rangle_{S3}\}\,,\\
|[31]_{FS}\rangle_{2}&=&\frac{1}{2}\{\sqrt{2}|[22]\rangle_{F1}|[31]\rangle_{S1}+
|[22]\rangle_{F1}|[31]\rangle_{S2}\nonumber\\
&&-|[22]\rangle_{F2}|[31]\rangle_{S3}\}\,,\\
|[31]_{FS}\rangle_{3}&=&\frac{1}{2}\{\sqrt{2}|[22]\rangle_{F2}|[31]\rangle_{S1}
-|[22]\rangle_{F2}|[31]\rangle_{S2}\nonumber\\
&&-|[22]\rangle_{F1}|[31]\rangle_{S3}\}\,,
\end{eqnarray}
the explicit forms of the flavor symmetry $[22]_{F}$ has been shown
in Ref. \cite{zou1}, and spin symmetry $[31]_{S}$ may be (for
$S_{z}=+1$)
\begin{eqnarray}
|[31]\rangle_{S1}&=&\frac{1}{\sqrt{12}}\{
3|\uparrow\uparrow\uparrow\downarrow\rangle
-|\uparrow\uparrow\downarrow\uparrow\rangle
-|\uparrow\downarrow\uparrow\uparrow\rangle\nonumber\\
&&-|\downarrow\uparrow\uparrow\uparrow\rangle\}\,,\\
|[31]\rangle_{S2}&=&\frac{1}{\sqrt{6}}\{2|\uparrow
\uparrow\downarrow\uparrow\rangle
-|\uparrow\downarrow\uparrow\uparrow\rangle
-|\downarrow\uparrow\uparrow\uparrow\rangle\}\,,\\
|[31]\rangle_{S3}&=&\frac{1}{\sqrt{2}}\{|\uparrow
\downarrow\uparrow\uparrow\rangle
-|\downarrow\uparrow\uparrow\uparrow\rangle\}\,,
\end{eqnarray}
(for $S_{z}=0$)
\begin{eqnarray}
|[31]\rangle_{S1}&=&\frac{1}{\sqrt{6}}\{
|\uparrow\uparrow\downarrow\downarrow\rangle
+|\downarrow\uparrow\uparrow\downarrow\rangle
+|\uparrow\downarrow\uparrow\downarrow\rangle
-|\downarrow\uparrow\downarrow\uparrow\rangle\nonumber\\
&&-|\uparrow\downarrow\downarrow\uparrow\rangle
-|\downarrow\downarrow\uparrow\uparrow\rangle\}\,,\\
|[31]\rangle_{S2}&=&\frac{1}{\sqrt{12}}\{
2|\uparrow\uparrow\downarrow\downarrow\rangle
-2|\downarrow\downarrow\uparrow\uparrow\rangle
+|\uparrow\downarrow\downarrow\uparrow\rangle
-|\downarrow\uparrow\uparrow\downarrow\rangle\nonumber\\
&&+|\downarrow\uparrow\downarrow\uparrow\rangle
-|\uparrow\downarrow\uparrow\downarrow\rangle\}\,,\\
|[31]\rangle_{S3}&=&-\frac{1}{2}\{|\downarrow
\uparrow\uparrow\downarrow\rangle
-|\uparrow\downarrow\downarrow\uparrow\rangle
+|\downarrow\uparrow\downarrow\uparrow\rangle
-|\uparrow\downarrow\uparrow\downarrow\rangle\}\,
\end{eqnarray}
\end{appendix}

%
%

%
%

%
%

%
%

%
%

%

%

\begin{thebibliography}{}
%
%

\bibitem{isgur} S. Capstick and N. Isgur, Phys. Rev. {\bf D34}  (1986)2809


\bibitem{glozman} L.Ya. Glozman and D. O. Riska, Phys. Rept. {\bf268}
(1996) 263

\bibitem{weise} N. Kaiser, P. Siegel and W. Weise, Phys. Lett. {\bf
B362} (1995) 23

\bibitem{oset} D. Jido et al., Nucl. Phys. {\bf A725} (2003) 181

\bibitem{oset2} D. Jido, M. D\"{o}ring and E. Oset, Phys. Rev. {\bf C77} (2008) 065207.

\bibitem{capstick} S. Capstick, B. D. Keister and D. Morel,
J. Phys. Conf. Ser.{\bf 69} (2007) 012016

\bibitem{riska} Q. B. Li and D. O. Riska, Nucl. Phys. {\bf
A766} (2006) 172

\bibitem{riska1} Q. B. Li and D. O. Riska, Phys. Rev. {\bf C73} (2006) 035201

\bibitem{riska2} Q. B. Li and D. O. Riska, Phys. Rev. {\bf C74} (2006) 015202

\bibitem{xie} J. J. Xie, B. S. Zou and H. C.
Chiang, Phys. Rev. {\bf C77} (2008) 015206

\bibitem{zou2}B. C. Liu and B. S. Zou , Phys. Rev. Lett.
{\bf96} (2006) 042002

\bibitem{qqq} S. Capstick and W. Roberts, Pro. Par. Nucl. Phys. {\bf
45} (2000) S241

\bibitem{qqq1} V. D. Burkert, Pro. Par. Nucl. Phys. {\bf 55} (2005) 108

\bibitem{helminen} C. Helminen and D. O. Riska, Nucl. Phys.
{\bf A699} (2002) 624

\bibitem{Kaiser} N. Kaiser, T. Waas and W. Weise, Nucl. Phys. {\bf
A612} (1997) 297

\bibitem{oset3}  M. D\"oring, E. Oset and B.S. Zou, Phys. Rev. {\bf C78} (2008) 025207

\bibitem{zou} B. S. Zou and D. O. Riska, Phys. Rev. Lett. {\bf 95}
(2005) 072001

\bibitem{zou1}C. S. An, D. O. Riska and B. S. Zou, Phys. Rev. {\bf C73} (2006)
035207

\bibitem{me} C. S. An, Q. B. Li, D. O. Riska and B. S. Zou,
Phys. Rev. {\bf C74} (2006) 055205, Erratum-ibid.{\bf C79} (2007)
069901

\bibitem{me1} C. S. An, Nucl. Phys. {\bf A797} (2007) 131 , Erratum-ibid.{\bf
A801} (2008) 82

\bibitem{charge} C. S. An and D. O. Riska, Eur. Phys. J. {\bf A37} (2008) 263

\bibitem{Lee} B. Juli\'{a}-D\'{\i}az, T.-S. H. Lee, T. Sato and
L. C. Smith, Phys. Rev. {\bf C75} (2007) 015205

\bibitem{chen}J. Q. Chen, \textit{Group Representation Theory for
Physicists,2nd edition}(World Scientific, Singapore, 1989)

\bibitem{Isgur2}R. Koniuk and N. Isgur, Phys. Rev. {\bf D21} (1980)
 1868, Erratum-ibid.{\bf
D23} (1981) 818

\bibitem{pdg}Particle Group Data, J. Phys.
{\bf G33} (2006) 1

\bibitem{prc} I. G. Aznauryan et al., Phys. Rev. {\bf C71} (2005)
 015201

\bibitem{clas} The CLAS Collaboration, Phys. Rev. Lett.
{\bf86} (2001) 1702

\bibitem{clasnew} The CLAS Collaboration, Phys. Rev. {\bf C76} (2007)
 015204


\end{thebibliography}
%

\end{document}